\documentclass[pra,showpacs,reprint,twocolumn,amsmath,amssymb,aps,superscriptaddress]{revtex4-1}

\usepackage{graphicx}
\usepackage{dcolumn}
\usepackage{bm}
\usepackage{epstopdf}
\usepackage{color}
\usepackage{times}
\usepackage{url}
\usepackage{tikz}
\usepackage{xspace}
\usepackage{xr}
\usepackage{chemformula}
\usepackage{float}

\newcommand{\dmit}{\ch{EtMe3Sb}\xspace}

\usetikzlibrary{decorations.markings}

\def\MarkLt{4pt}
\def\MarkSep{2pt}

\tikzset{
  TwoMarks/.style={
    postaction={decorate,
      decoration={
        markings,
        mark=at position #1 with
          {
              \begin{scope}[xslant=0.2]
              \draw[line width=\MarkSep,white,-] (0pt,-\MarkLt) -- (0pt,\MarkLt) ;
              \draw[-] (-0.5*\MarkSep,-\MarkLt) -- (-0.5*\MarkSep,\MarkLt) ;
              \draw[-] (0.5*\MarkSep,-\MarkLt) -- (0.5*\MarkSep,\MarkLt) ;
              \end{scope}
          }
       }
    }
  },
  TwoMarks/.default={0.5},
}

\def\zdir{\tikz[baseline=-1ex]{
\fill (0,0) circle (1.5pt) coordinate (A);
\fill (0,2ex) circle (1.5pt) coordinate (B);
\draw[TwoMarks=0.5] (A)--(B);}}

\def\adir{\tikz[baseline=-1ex]{
\fill (0,0) circle (1.5pt) coordinate (A);
\fill (2ex,0) circle (1.5pt) coordinate (B);
\draw (A)--(B);}}

\def\bdir{\tikz[baseline=-1ex]{
\fill (0,-1.6ex) circle (1.5pt) coordinate (A);
\fill (0.6ex,0) circle (1.5pt) coordinate (B);
\draw [densely dashed] (A)--(B);}}

\def\cdir{\tikz[baseline=-1ex]{
\fill (0,0) circle (1.5pt) coordinate (A);
\fill (1.4ex,-1.6ex) circle (1.5pt) coordinate (B);
\draw [densely dotted] (A)--(B);}}

\def\ringc{\tikz[baseline=-1ex]{
\fill (0,0) circle (1.5pt) coordinate (A);
\fill (2ex,0) circle (1.5pt) coordinate (B);
\fill (1.4ex,-1.6ex) circle (1.5pt) coordinate (C);
\fill (-0.6ex,-1.6ex) circle (1.5pt) coordinate (D);
\draw (A)--(B);
\draw (C)--(D);
\draw [densely dashed] (A)--(D);
\draw [densely dashed] (B)--(C);
\draw [densely dotted] (A)--(C);}}

\def\ringb{\tikz[baseline=-1ex]{
\fill (0,0) circle (1.5pt) coordinate (A);
\fill (2ex,0) circle (1.5pt) coordinate (B);
\fill (3.4ex,-1.6ex) circle (1.5pt) coordinate (C);
\fill (1.4ex,-1.6ex) circle (1.5pt) coordinate (D);
\draw (A)--(B);
\draw (C)--(D);
\draw [densely dotted] (A)--(D);
\draw [densely dotted] (B)--(C);
\draw [densely dashed] (B)--(D);}}

\def\ringa{\tikz[baseline=-1ex]{
\fill (0,0) circle (1.5pt) coordinate (A);
\fill (1.4ex,-1.6ex) circle (1.5pt) coordinate (B);
\fill (0.8ex,-3.2ex) circle (1.5pt) coordinate (C);
\fill (-0.6ex,-1.6ex) circle (1.5pt) coordinate (D);
\draw [densely dotted] (A)--(B);
\draw [densely dotted] (C)--(D);
\draw [densely dashed] (A)--(D);
\draw [densely dashed] (B)--(C);
\draw (B)--(D);}}

\def\ringthree{\tikz[baseline=-1ex]{
\fill (0,0) circle (1.5pt) coordinate (A);
\fill (2ex,0) circle (1.5pt) coordinate (B);
\fill (1.4ex,-1.6ex) circle (1.5pt) coordinate (C);
\draw (A)--(B);
\draw [densely dotted] (A)--(C);
\draw [densely dashed] (B)--(C);}}

\begin{document}
	
\title{Frustration, ring exchange, and the absence of long-range order in EtMe$_3$Sb[Pd(dmit)$_2$]$_2$: from first principles to many-body theory}
\author{E.~P.~Kenny}
\email{elisekenny@gmail.com}
\affiliation{School of Mathematics and Physics, The University of Queensland, Brisbane, Queensland, Australia}
\author{G.~David}
\altaffiliation{Current address: School of Chemistry, University of Nottingham, University Park, Nottingham NG7 2RD, United Kingdom} 
\affiliation{Aix-Marseille Univ, CNRS, ICR, Marseille, France}
\author{N.~Ferr\'e}
\affiliation{Aix-Marseille Univ, CNRS, ICR, Marseille, France}
\author{A.~C.~Jacko}
\affiliation{School of Mathematics and Physics, The University of Queensland, Brisbane, Queensland, Australia}
\author{B.~J.~Powell}
\affiliation{School of Mathematics and Physics, The University of Queensland, Brisbane, Queensland, Australia}

\date{\today}

\begin{abstract}
We parameterize Hubbard and  spin models for EtMe$_3$Sb[Pd(dmit)$_2$]$_2$ from broken symmetry density functional calculations. This gives a scalene triangular model where the largest net exchange interaction is three times larger than the mean interchain coupling. The chain random phase approximation  shows that the difference in the interchain couplings is equivalent to a bipartite interchain coupling, favoring long-range magnetic order. This competes with ring exchange, which favors quantum disorder. Ring exchange wins. We predict that the thermal conductivity, $\kappa$, along the chain direction is much larger than that along the crystallographic axes and that $\kappa/T\rightarrow0$ as $T\rightarrow0$ along the crystallographic axes, but that  $\kappa/T\rightarrow\textrm{a constant}>0$ as $T\rightarrow0$ along the chain direction.
\end{abstract}

\maketitle

\section{Introduction}

EtMe$_3$Sb[Pd(dmit)$_2$]$_2$ (\dmit) is a quantum spin liquid (QSL) candidate shrouded in mystery. It lacks magnetic ordering  down to the lowest temperatures measured \cite{Itou2010, Itou2011, Itou2008}, but the physics that results in a quantum disordered state remains under debate. \dmit shares important structural motifs with the quantum spin liquids $\kappa$-(BEDT-TTF)$_2$Cu$_2$(CN)$_3$ ($\kappa$-Cu) and $\kappa$-(BEDT-TTF)$_2$Ag$_2$(CN)$_3$  ($\kappa$-Ag). A crucial question is: how closely related are their ground states?

\begin{figure}
	\centering
	\includegraphics[width=\columnwidth]{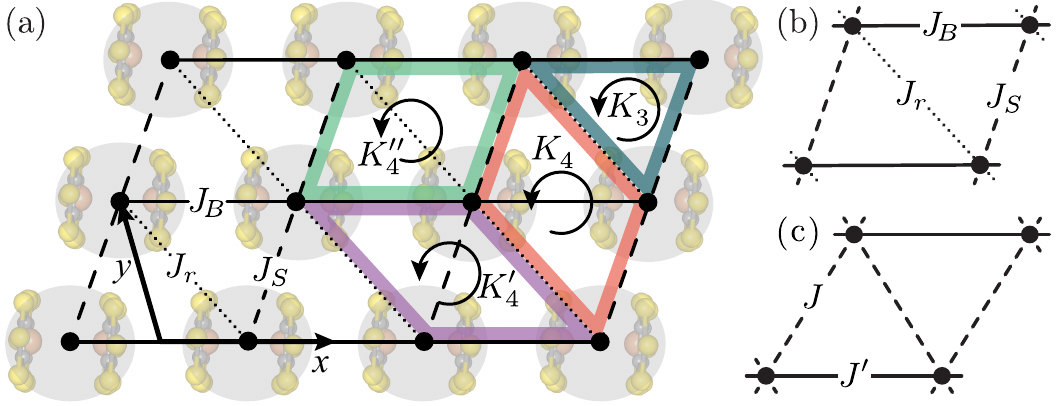}
	\caption{We find three significant antiferromagnetic nearest-neighbor couplings within the layers of [Pd(dmit)$_2$]$_2$ dimers (a). The largest, $J_B$, is in the dimer stacking direction. The others, $J_r\ne J_S$ lead to a scalene lattice (b). We show that the physics of this model differs importantly from the isosceles triangular lattice relevant to the closely related BEDT-TTF salts (c). We also calculate the three and four-site ring exchange interactions, $K_3$, $K_4$, $K^\prime_4$, and $K^{\prime\prime}_4$ (a). The interlayer coupling, $J_z$, is approximately perpendicular to the page.}
	\label{fig:geometry}
\end{figure}

\dmit, $\kappa$-Cu, and $\kappa$-Ag all form structures with alternating layers of organic molecules and counter-ions. In all three materials, the organic molecules dimerize with one unpaired electron found on each dimer in the insulating phase. However, the spacial arrangement the dimers differs. Within $\kappa$-Cu and $\kappa$-Ag, neighboring dimers are almost perpendicular to one another, whereas in \dmit, the dimers  (gray circles in Fig. \ref{fig:geometry}a) form quasi-one-dimensional stacks (along the horizontal in Fig. \ref{fig:geometry}a).

$\kappa$-Cu and $\kappa$-Ag are Mott insulators. In the strong coupling limit, where the Hubbard $U$ is much greater than the largest interdimer hopping integral, $t$, their insulating phase is described by the isosceles triangular Heisenberg model (Fig. \ref{fig:geometry}c). This model has two candidate QSL phases. 
 Firstly, a QSL has been suggested in the region $0.6\lesssim J^\prime/J\lesssim 0.9$ \cite{Scriven2012,Bishop2009,Zheng1999}, for which the ground state remains controversial.
 Secondly, the large $J'/J$ limit is adiabatically connected to the Tomonaga-Luttinger liquid (TLL) expected for uncoupled chains, $J^\prime/J\gtrsim 1.4$ \cite{Zheng1999, Zheng2007, Yunoki2006, Hayashi2007, Tocchio2014, Ghorbani2016}. Theories in this regime show an emergent  `one-dimensionalization' whereby the many-body state is more one-dimensional than the underlying Hamiltonian  \cite{Balents2010,Kohno2007,Starykh2010,Powell2007}.  
However, the validity of the strong coupling limit in these materials is uncertain because both undergo Mott metal-insulator transitions under moderate pressures. This motivates the inclusion of higher order terms in the spin model, most importantly ring exchange. It has been shown that these can also cause QSL phases \cite{Motrunich2005,Holt2014,Merino2014,Misguich1998, Misguich1999}.

Many early studies explored the possibility that the spin liquid in \dmit can be explained by one of the above theories. 
However, the lower symmetry of \dmit means that all three exchange interactions are different, \textit{i.e.}, it is described by a scalene triangular lattice, Fig. \ref{fig:geometry}a,b. \dmit is also close to a Mott transition and so ring exchange is likely to be important.

In this work we parametrize the spin model of \dmit including the scalene Heisenberg and ring exchange interactions using broken symmetry density functional theory (BS-DFT) calculations \cite{assfeld1996, David2017, Coulaud2013, Coulaud2012}. We find that the strongest exchange coupling is along the dimer stacking direction ($J_B$; \textit{cf}. Fig. \ref{fig:geometry}a). We solve our model via the chain random phase approximation (CRPA) around the large $J_B$ limit. In this approach, one starts from the exact form for the one-dimensional magnetic susceptibility of a Heisenberg spin-1/2 chain and treats interchain interactions via the RPA \cite{Schulz1996,Bocquet2001}. On an isosceles triangular lattice, the interchain interactions are perfectly frustrated. Within the CRPA, this prevents ordering at any temperature \cite{Bocquet2001, Kenny2019}. In \dmit, we find that the anisotropy in the interchain coupling leads to an effective unfrustrated interchain interaction, given by the difference of the interchain couplings ($\delta J_y=J_r\!-\!J_S$). This favors long-range order. On the other hand, ring exchange favors quantum disorder \cite{Motrunich2005, Misguich1998, Misguich1999, Merino2014, Holt2014}.
Combining our BS-DFT and CRPA results shows that the absence of long-range magnetic order in \dmit springs from the interplay of one-dimensionalization and ring exchange, leading us to propose that the ground state of \dmit is adiabatically connected to the TLL.

\section{Parametrization of Spin Model with Broken-Symmetry Density Functional Theory}

The low-energy physics of the insulating phase of \dmit is described by an extended Hubbard model \cite{Powell2011,Kanoda2011,Nakamura2012},
\begin{widetext}
\begin{equation}
\mathcal{H}_\mathrm{Hubbard}=
\sum_{ij\sigma}t_{ij}c^\dagger_{i\sigma}c^{ }_{j\sigma} + U\sum_{i}n_{i\uparrow}n_{i\downarrow} 
+\frac{1}{2}\sum_{i\neq j} \Bigg[\left( \sum_{\sigma\rho}V_{ij}^*c^\dagger_{i\sigma}c^\dagger_{j\rho}c^{ }_{j\rho}c^{ }_{i\sigma} \right) 
\label{eq:hubb} 
+
(J_{ij}^\mathrm{DE} + J_{ij}^\mathrm{SP}){\bm S}_{i} \cdot {\bm S}_{j} 
- \frac{J_{ij}^\mathrm{DE}}{2} {c}_{i\uparrow}^\dagger {c}_{i\downarrow}^\dagger  {c}_{j\uparrow} {c}_{j\downarrow}\Bigg], 
\end{equation}
where $c^\dagger_{i\sigma}$ ($c^{ }_{i\sigma}$) creates (destroys) an electron with spin $\sigma$ on site (dimer) $i$, $t_{ij}$ is the hopping between sites, $U$ is the effective on-site Coulomb repulsion, $n_{i\sigma}=c^\dagger_{i\sigma}c^{ }_{i\sigma}$ is the electron density, $V_{ij}^*=V_{ij}+(J_{ij}^\mathrm{DE}/4)$, $V_{ij}$ is the Coulomb repulsion between electrons on different dimers, $J_{ij}^\mathrm{DE}$ is the interdimer direct exchange, and $J_{ij}^\mathrm{SP}$ is the interdimer spin polarization. While there have been a number of calculations of $t_{ij}$ \cite{Scriven2012,Jacko2013,Tsumuraya2013,Sup},  only \citeauthor{Nakamura2012} \cite{Nakamura2012} have previously estimated $U$ and $V_{ij}$. They also calculated the direct exchange $J_{ij}^\mathrm{DE}$.  Although $|J_{ij}^\mathrm{DE}|\ll U, V_{ij}$, \citeauthor{Nakamura2012}'s parameters show that $J_{ij}^\mathrm{DE}$ is non-negligible on the scale of the superexchange interaction, $J_{ij}^\mathrm{SE}$ \footnote{Our $J_{ij}^\mathrm{DE}=-2J_{ij}$ in \citeauthor{Nakamura2012}'s notation.}. 

We construct an effective low-energy spin model of the Mott insulating phase for $t_{B}\ll U-V_B$.
As well as the usual superexchange interactions, we also retain the three- and four-site ring exchange, illustrated in Fig. \ref{fig:geometry}a,
\begin{align}
\mathcal{H} = &
\frac{J_B}{2}\sum_{\adir}\hat{P}_{ij} + \frac{J_z}{2}\sum_{\zdir}\hat{P}_{ij}
+ \frac{J_S}{2}\sum_{\bdir}\hat{P}_{ij}
+ \frac{J_r}{2}\sum_{\cdir}\hat{P}_{ij}\nonumber\\&
+ \frac{K_3}{2}\sum_{\ringthree}\left(\hat{P}_{ijk}+\hat{P}_{lkj}\right)
+ \frac{K_4}{2}\sum_{\ringa}\left(\hat{P}_{ijkl}+\hat{P}_{lkji}\right)
+ \frac{K^{\prime}_4}{2}\sum_{\ringb}\left(\hat{P}_{ijkl}+\hat{P}_{lkji}\right)
+ \frac{K^{\prime\prime}_4}{2}\sum_{\ringc}\left(\hat{P}_{ijkl}+\hat{P}_{lkji}\right)\label{eq:ring_ham}
\end{align} 
\end{widetext}
where 
$J_{ij}^\mathrm{SE}=4t_{ij}^2/U_{ij}$, 
$U_{ij}=U-V_{ij}$,
$J_{ij}=J_{ij}^\mathrm{SE} + J_{ij}^\mathrm{DE} + J_{ij}^\mathrm{SP}$ [we retain only $\{i,j\}\in\{B,S,r\}$ (\textit{cf}. Fig. \ref{fig:geometry}a,b)],
$K_3=3t_Bt_rt_S/U_\textrm{eff}^{2}$, $K_4\!=\!80t_{r}^2t_{S}^2/U_\textrm{eff}^3$, 
$K_4^\prime\!=\!80t_{r}^2t_{B}^2/U_\textrm{eff}^3$, 
$K_4^{\prime\prime}\!=\!80t_{S}^2t_{B}^2/U_\textrm{eff}^3$, 
$U_\textrm{eff}\simeq U-\frac13(V_B+V_S+V_r)$, which is a reasonable approximation as the three $V$'s do not vary greatly, \textit{vida infra},
$\hat{P}_{ij}\!=\!2\bm{S}_i\cdot\bm{S}_j+\frac{1}{2}$, $\hat{P}_{ijk}\!=\!\hat{P}_{ij}\hat{P}_{jk}$, 
$\hat{P}_{ijkl}\!=\!\hat{P}_{ij}\hat{P}_{jk}\hat{P}_{kl}$,
and $\hat{P}_{ijk}$ and $\hat{P}_{ijkl}$ cyclically permute spin states around the plaquettes shown (with dashing to match Figs. \ref{fig:geometry}a,b). 

Significant effort has been expended parameterizing tight-binding models for \dmit from DFT \cite{Scriven2012, Tsumuraya2013, Jacko2013, Sup}. However, these calculations do not give a direct parameterization of the spin model [Eq. (\ref{eq:ring_ham})] because they do not enable the calculation of $U$, $V_{ij}$, $J_{ij}^\mathrm{DE}$, or $J_{ij}^\mathrm{SP}$. \citeauthor{Nakamura2012} \cite{Nakamura2012} addressed this by performing constrained RPA calculations, which do provide estimates of the Coulomb interactions. 
All these calculations are based on pure density functionals, \textit{i.e.}, the local density approximation (LDA) or generalized gradient approximations (GGA), which are known to perform poorly for parametrizing magnetic interactions \cite{Malrieu2014,Moreira2002,Zein2009,Phillips2012}. In particular, they underestimate $J_{ij}^\mathrm{SE}$ \cite{Coulaud2012}. For further discussion of functionals, see the Supplementary Material \cite{Sup}.

LDA+U calculations are not straightforward in these molecular systems; like many inorganic and organometallic magnets, the spins are delocalized over a dimer rather than being centered on a single atom. However, hybrid functionals have been shown to provide similar accuracy as the LDA+U calculations in many molecular systems \cite{Malrieu2014,Rivero2009}. 

Previous tight-binding models based on either the monomer or dimer models of \dmit \cite{Scriven2012, Tsumuraya2013, Jacko2013, Nakamura2012, Powell2011}  neglect states outside of a small energy window near the Fermi surface. These models are based on Kohn-Sham eigenvalues, which have no formal correspondence to nature but are a device for calculating the total density \cite{KohnSham1965}. In practice, Kohn-Sham eigenvalues poorly reproduce energy differences  even in weakly correlated materials \cite{Jones1989, Perdew1985} and dramatically fail in strongly correlated materials. For example, in \dmit, $\kappa$-Cu, and $\kappa$-Ag the Kohn-Sham band structure is metallic \cite{Scriven2012, Tsumuraya2013, Jacko2013, Nakamura2012, Powell2011} rather than insulating, as in experiment \cite{Powell2011,Kanoda2011}.

In BS-DFT, one calculates exchange interactions by comparing total energies of the full atomistic Hamiltonian, which have a formal basis in DFT and are highly accurate in practice. Recent advances \cite{David2017, Coulaud2013, Coulaud2012} have made it possible to isolate distinct physical contributions to the total exchange and even the Hubbard parameters from this approach. Thus, \dmit provides a valuable opportunity for a comparison of BS-DFT with constrained RPA. BS-DFT calculations are based on a cluster, rather than an infinite crystal. This is a double-edged sword. Finite size effects need to considered, but the finite size makes hybrid functionals, which include exact exchange interactions, computationally tractable.

In light of these considerations, we calculated $J_{ij}^\mathrm{SE}$, $J_{ij}^\mathrm{DE}$,  $J_{ij}^\mathrm{SP}$,  $t_{ij}$, and $U-V_{ij}$ for each nearest neighbor pair of dimers from a series of BS-DFT calculations. 
We  utilize the frozen orbital capabilities of the local self-consistent field method \cite{David2017, Coulaud2013, Coulaud2012, assfeld1996}.
We use the ``quasi-restricted" orbital (QRO) approach \cite{Neese2006} with LANDL2DZ effective core potential and basis set for palladium and antimony \cite{hay1985b,wadt1985a} and 6-31+G* basis set \cite{clark1983a,ditchfield1971a,hariharan1973a,hehre1972a} for other atoms and with hybrid B3LYP functional \cite{B3LYPa} in ORCA \cite{ORCA}.
See the Supplementary Material \cite{Sup} for results with a different amount of Hartree-Fock exchange, which has been shown to have an effect on BS-DFT couplings \cite{David2017, Martin1997, Sorkin2008, Reiher2001, Swart2004, Dai2005, Lawson2005, Pierloot2006, Rong2006, Brewer2006, Vargas2006, Ovcharenko2007, Valero2008}. We included the six nearest cations to each Pd(dmit)$_2$ tetramer; benchmarking revealed that the calculated exchange interactions are well converged at this cluster size. We use the experimental crystal structure measured at 4\,K \cite{Yugo}.

\begin{figure}
	\centering
	\includegraphics[width=0.95\columnwidth]{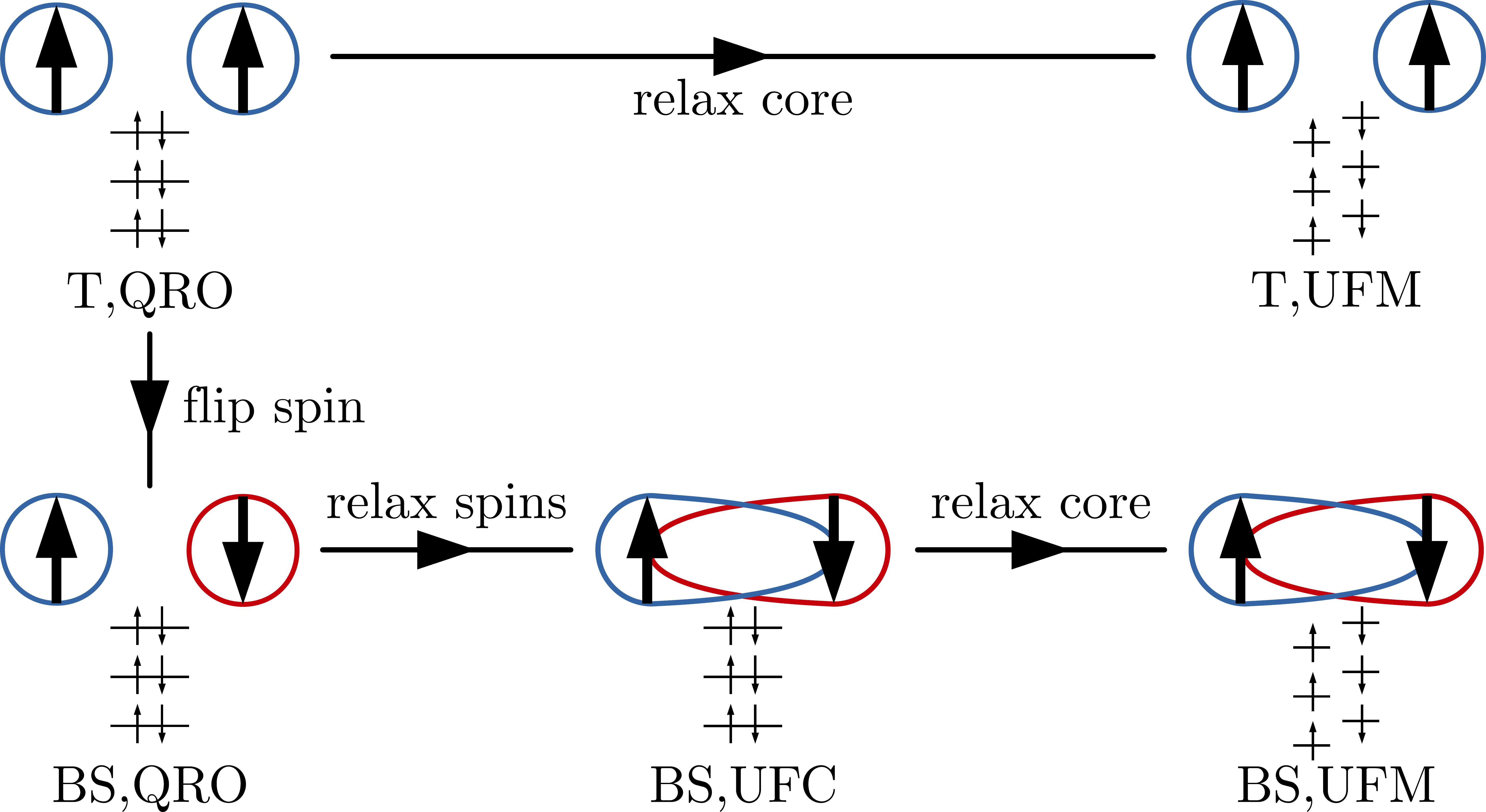}
	\caption{Illustration of the decomposition of the magnetic exchange coupling calculation.  Blue and red circles represent magnetic orbitals (singly) occupied by spin-up and spin-down electrons respectively. Horizontal black lines and vertical arrows  represent non-magnetic occupied orbitals (also known as core orbitals).}
	\label{fig:DecompoPath}
\end{figure}

As illustrated in Fig \ref{fig:DecompoPath}, we start with a triplet state in the quasi-restricted open-shell formalism (T,QRO). We split the high spin dimer one-electron orbitals into two different sets, (i) the two same spin localized magnetic orbitals and (ii) the remaining (non-magnetic) ones.
A BS solution is found by flipping the individual spin state of one magnetic orbital (BS,QRO in Fig \ref{fig:DecompoPath}), giving us the direct exchange, $J_{ij}^\mathrm{DE}$.
We then relax (i.e. delocalize) only the magnetic orbitals in the BS solution (BS,UFC in Fig \ref{fig:DecompoPath}), which allows us to calculate the superexchange $J_{ij}^\mathrm{SE}$ and the Hubbard parameters $t_{ij}$ and $U-V_{ij}$, as described in \cite{David2017}.
The magnetic orbitals are then kept frozen in both the triplet and BS states while the non-magnetic ones are relaxed (T,UFM and BS,UFM in Fig \ref{fig:DecompoPath}), eventually giving the spin-polarization contribution, $J_{ij}^\mathrm{SP}$.

Our BS-DFT results are shown in Table \ref{tab:dft}. We calculate small values for the spin-polarisation contribution, $J^\mathrm{SP}\sim 0.05J_B$, and henceforth neglect this term. The unfrustrated interlayer coupling is $J_z=0.06$\,K.  

Our values for the total Heisenberg exchange ($J_{ij}$) between dimers reveal that the exchange along the dimer stacking direction, $J_B$ (see Fig. \ref{fig:geometry}a), is significantly larger than the others; $J_r/J_B=0.35$ and $J_S/J_B=0.30$. In what follows, it will be convenient to make a  change of variables to the average of the interchain couplings, $\bar{J_y}=\frac{1}{2}\left(J_S\!+\!J_r\right)\!=\!135$\,K, and their difference, $\delta J_y=J_r\!-\!J_S\!=\!22$\,K. 

\strut

\begin{table}
	\centering
	\begin{tabular}{c|ccccc}
		$ij$& $J_{ij}$\,(K) & $J_{ij}^\mathrm{SE}$\,(K) & $J_{ij}^\mathrm{DE}$\,(K) 
		& $t_{ij}$\,(meV) & $U\!-\!V_{ij}$\,(meV)\\ 
		\hline 
		\hline
		B & 414 & 481 & -67 & 80 & 565\\ 
		r & 146 & 168 & -22 & 47 & 600\\
		S & 124 & 168 & -44 & 44 & 519\\
	\end{tabular}
	\caption{Calculated nearest neighbour interactions between dimers of \dmit, as shown in Figure \ref{fig:geometry}a. The full Heisenberg exchange ($J_{ij}$) is a sum of the superexchange ($J_\mathrm{SE}$) and the direct exchange ($J_\mathrm{DE}$). $t_{ij}$ is the effective hopping between dimers and $U\!-\!V_{ij}$ is the effective Coulomb interaction on each dimer.}\label{tab:dft}
\end{table}

In the limit $\delta J_y$ and $J_z\rightarrow 0$, the lattice becomes a quasi-one-dimensional isosceles model (\textit{cf}. Fig. \ref{fig:geometry}c). Numerical studies have shown that this model remains quasi-one-dimensional for $\bar{J_y}/J_B<0.7$  \cite{Weng2006, Yunoki2006, Hayashi2007, Pardini2008, Jiang2009, Heidarian2009, Tay2010}.  The unfrustrated limit (explored by \citeauthor{Schulz1996} \cite{Schulz1996}) occurs when $J_S\rightarrow 0$ or $J_r\rightarrow 0$, in which case the magnitude of the unfrustrated interchain coupling is $|\delta J_y|$. This model is quasi-one-dimensional for $(|\delta J_y|\!+\!|J_z|)/J_B<0.3$  \cite{Yasuda2005}.
In \dmit, both the frustrated component of the interchain coupling, $\bar{J_y}/J_B\!=\!0.33\ll0.7$, and the total unfrustrated component, $\left(|\delta J_y|\!+\!|J_z|\right)/J_B\!=\!0.05\ll0.3$, are comfortably within quasi-one-dimensional limits. 

We determine the ring exchange parameters (cf. Eq. \ref{eq:ring_ham}) using our values of $t_{ij}$ and $U\!-\!V_{ij}$. The three-site ring exchange, $K_3\!=\!18$\,K, simply renormalizes the Heisenberg couplings within the Pd(dmit)$_2$ planes; $J_B\rightarrow J_B\!+\!K_3$ and $\bar{J_y}\rightarrow\bar{J_y}\!+\!K_3$ \cite{Misguich1998}. The four-site terms are slightly larger: 
$K_4\!=\!23$\,K, 
$K_4^\prime\!=\!76$\,K, 
$K_4^{\prime\prime}\!=\!66$\,K. 
These terms are also more consequential due to effective interactions in additional directions within the lattice. To include them in an effective Heisenberg model, we use a leading order mean-field approximation \cite{Holt2014}, $\langle \bm{S}_\alpha\cdot\bm{S}_\beta\rangle = S^2\cos(\bm{k}\cdot \bm{r}_{\alpha\beta})$, where $\bm{r}_{\alpha\beta}$ is the vector from site $\alpha$ to site $\beta$. This results in renormalised exchange couplings $J_{ij}$ in the $\bm x$, $\pm\frac{1}{2}\bm{x}-\bm{y}$,  $\pm\frac{3}{2}\bm{x}-\bm{y}$, and $2\bm{y}$ directions (see axes in Fig. \ref{fig:geometry}a). 

\section{$\textrm{EtMe$_3$Sb}$ as a Quasi-1D Spin System}

\subsection{N\'eel Ordering Temperature}

The N\'eel ordering temperature, $T_N$, of a lattice of weakly coupled chains can be calculated using the CRPA expression for the three-dimensional dynamical magnetic susceptibility, \cite{Scalapino1975, Schulz1996, Essler1997, Bocquet2001}
\begin{equation}
\chi\left(\omega,\bm{k}, t\right)=\frac{\chi_\mathrm{chain}(\omega,k_x, t)}{1-2\tilde{J}_\perp\left(\bm k\right)\chi_\mathrm{chain}(\omega,k_x, t)},
\end{equation}
where $t\!=\!k_B T/(J_B+K_3)$ is the reduced temperature, $\tilde{J}_{\perp}(\bm k)$ is the Fourier transform of the interchain coupling and $\bm{k}\!=\!\left(k_x,k_y,k_z\right)$ is the crystal momentum along the axes in Fig. \ref{fig:geometry}a in units of the inverse lattice spacing. The dynamical susceptibility for a single Heisenberg chain, calculated from a combination of the Bethe ansatz and field theory techniques, is \cite{Bethe1931,  Schulz1983, Schulz1986,  Barzykin2000, TsvelikBook} 
\begin{widetext}
\begin{equation}
	\chi_\mathrm{chain}(\omega,k_0, t) 
	=-\frac{\sqrt{\ln({\Lambda}/{t})}}{2t(2\pi)^{3/2}}
	\frac{\Gamma\left(\frac{1}{4}-i\frac{\omega-u k_0}{4\pi t}\right)}{\Gamma\left(\frac{3}{4}-i\frac{\omega-u k_0}{4\pi t}\right)}\frac{\Gamma\left(\frac{1}{4}-i\frac{\omega+u k_0}{4\pi t}\right)}{\Gamma\left(\frac{3}{4}-i\frac{\omega+u k_0}{4\pi t}\right)},
	\label{eq:chain_chi}
\end{equation}
\end{widetext}
where $k_0=k_x-\pi$, $\Gamma(x)$ is the Euler gamma function, $u=\frac{\pi}{2}J_{x}b_0$ is the spin velocity, $b_0$ is the interdimer separation along the quasi-one-dimensional stack, 
and $\Lambda\simeq24.27$  \cite{Barzykin2001}. 
The N\'eel temperature, $T_N$, corresponds to the zero frequency pole in $\chi\left(\omega,\bm{k}, t\right)$ when $2\tilde{J}_{\perp}\left(\bm{k}\right)\chi_\mathrm{chain}(0,k_x)|_{T=T_N}=1$. The instability occurs at the maximum of $J_{\perp}\left(\bm{k}\right)\chi_\mathrm{chain}(0,k_x)$. 

\begin{figure}
	\centering
	\includegraphics[width=0.95\columnwidth]{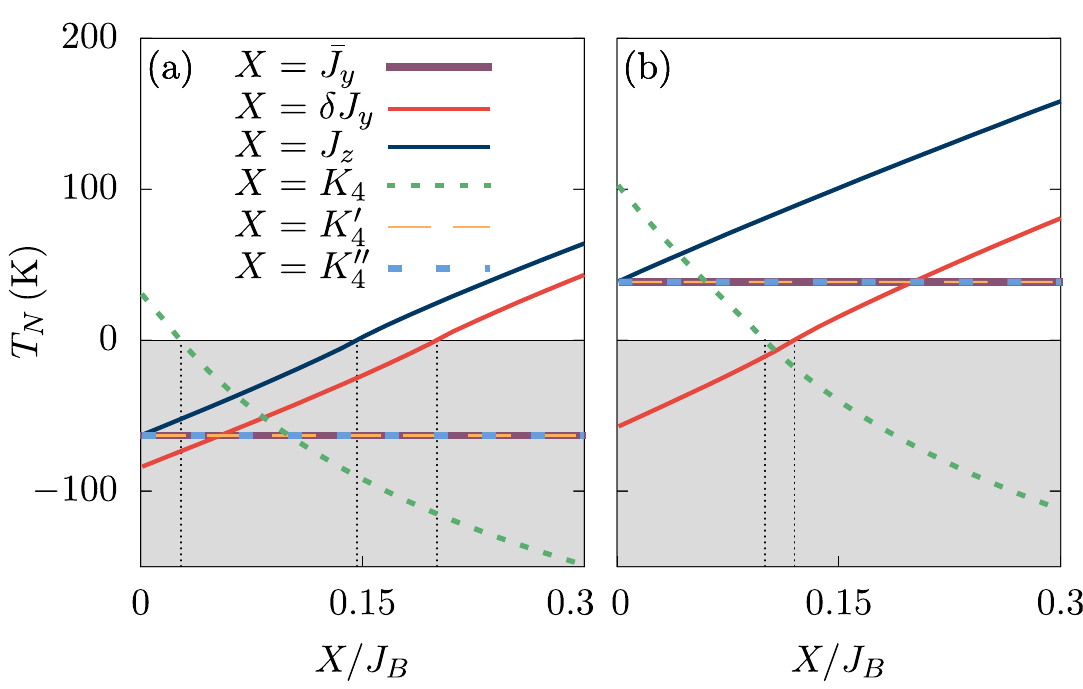}
	\caption{Numerical calculation of the ordering temperature as a function of each parameter in Eq. (\ref{eq:ring_ham}). Each line corresponds to varying one parameter, $X$, and keeping the others constant. In (a), all parameters (except for $X$) are set to our values for \dmit. In this case, solutions for $T_N$ are only positive when $K_4$ is very small or $\delta J_y$, $J_z$ are large. Solutions for $T_N$ become negative and complex when $2K_4>|\delta J_y|+|J_z|$ (for these unphysical solutions only the real part is shown). The vertical black lines indicate the parameter values for which this transition occurs. In (b), we set $K_4=0.06J_B$ and $\delta J_y=0.2J_B$ as an example of the regime $2K_4<|\delta J_y|+|J_z|$. In both cases, the solution is exactly reproduced with Eq. (\ref{eq:Tn_ring}); it is independent of the magnitudes of $\bar{J_y}$, $K_4^\prime$, and $K_4^{\prime\prime}$}
	\label{fig:num_explore}
\end{figure}

\begin{figure}
	\centering
	\includegraphics[width=0.95\columnwidth]{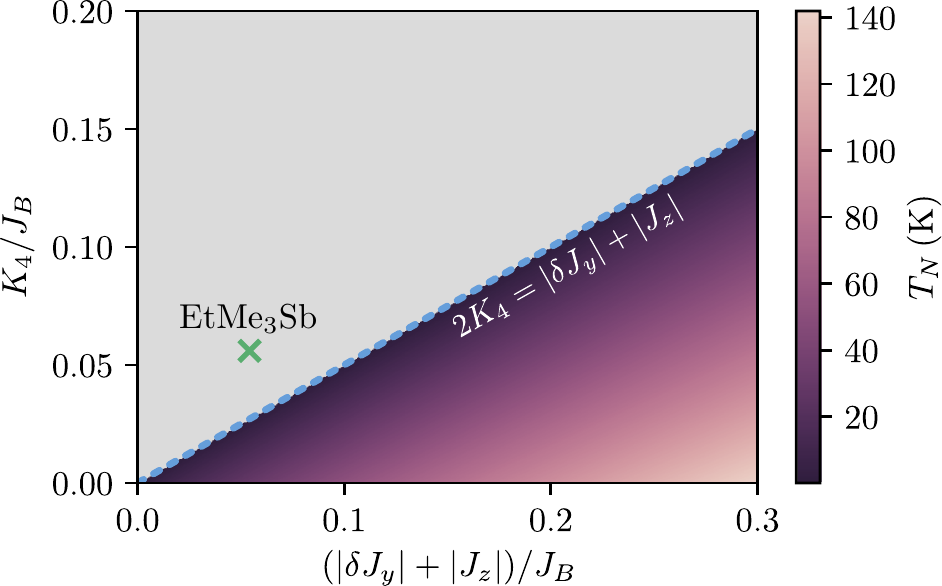}
	\caption{Numerical calculation of the ordering temperature as a function of the interchain ring exchange, $K_4$, and the unfrustrated interchain couplings, $\delta J_y$ and $J_z$. Our calculated parameters for \dmit are marked by the green cross. We have used our calculated values for the other couplings in Eq. (\ref{eq:ring_ham}); $\bar{J}_y\!=\!135$\,K, $K_3=18$\,K, $K^\prime_4\!=\!76$\,K, and $K^{\prime\prime}_4\!=\!66$\,K. The gray area indicates where there is no real positive solution for $T_N$. This occurs when $2K_4>|\delta J_y|+|J_z|$ (as shown with the blue dashed line). For all points on this graph, $k_0$ and $k_y+\pi/2$ are both less than $10^{-4}$.}
	\label{fig:Tn_plot}
\end{figure}

Numerical exploration of our system (see Fig. \ref{fig:num_explore}) reveals that $T_N$ is affected only negligibly by $J_y$, $K^{\prime}$, $K^{\prime\prime}$, but significantly by $\delta J_y$, $J_z$, $K_4$. We find positive, real solutions only when $2K_4<|\delta J_y|+|J_z|$. We also find, analytically, that only the magnitudes (and not the signs) of each interaction affect $T_N$. In light of this, Fig. \ref{fig:Tn_plot} shows a numerical calculation of $T_N$ as a function of the unfrustrated couplings and $K_4$. In the gray regions of Figs.  \ref{fig:num_explore} and \ref{fig:Tn_plot}, where $2K_4<|\delta J_y|+|J_z|$, there is no real positive solution. This implies that there is no long-range magnetic order, and that this state is in the same phase as the TLL.

For all points in Fig. \ref{fig:Tn_plot} (including the gray zone) we find $k_0\approx0$ and $k_y\approx-\pi/2$. Taking the limits $k_0\rightarrow0$ and $k_y\rightarrow-\pi/2$ analytically leads to
\begin{equation}
\frac{T_{N}}{J_B}=0.556\frac{|\delta J_y| + |J_z| - 2K_4}{J_B+3K_4}\sqrt{\log\left(\frac{\Lambda}{T_{N}}\right)}.\label{eq:Tn_ring}
\end{equation}
Thus, for $K_4=0$ one finds that $T_N \sim |\delta J_y| + |J_z|$. This agrees precisely with the prediction for a cuboidal model with chains along the $x$-axis coupled by bipartite exchange interactions $\delta J_y$, along the $y$-axis and $J_z$ along the $z$-axis \cite{Schulz1996}. We find that Figs. \ref{fig:num_explore} and \ref{fig:Tn_plot} are perfectly reproduced by Eq. (\ref{eq:Tn_ring}), confirming the relevance of this limit. Thus, we deduce three things: $\bar{J}_y$ is the frustrated part of the in-plane interchain interaction, which does not lead to long-range magnetic order and has no influence on $T_N$ (for $\delta J_y \neq 0$). $\delta J_y$ is the unfrustrated part of the in-plane interchain interaction, which can drive a magnetic instability and has a strong influence on $T_N$. $K_4\ne0$ strongly suppresses magnetic order (as $K_4\propto t_r^2t_S^2$ and therefore cannot be negative). 

Our DFT parameterization yields $2K_4=46$\,K, which is greater than our value for $|\delta J_y|+|J_z|=21$\,K. The solution to Eq. \ref{eq:Tn_ring} for \dmit is then unphysical; our quasi-1D model, with four-membered ring exchange, predicts that \dmit will not order magnetically down to $T=0$. This agrees with experiments, where ordering is not detected down to 19.4\,mK \cite{Itou2010}.

\subsection{Experimental Predictions}

In light of our findings, we propose that the `spin-liquid' behaviour in \dmit is a remnant of the TLL found in an isolated chain, similar to the state observed above $T_N=0.6$\,K in Cs$_2$CuCl$_4$ \cite{Kohno2007, Coldea1997}. This provides a natural explanation for the observed low temperature behaviour in \dmit. The  heat capacity \cite{Yamashita2011} reveals gapless excitiations from the ground state. This is consistent with the gapless spinon excitations expected in a TLL \cite{GiamachiBook}. The $^{13}$C nuclear spin-lattice relaxation rate shows a peak at 1\,K \cite{Itou2010}. We propose that this could be explained by short range correlations caused by the unfrustrated couplings ($\delta J_y$ and $J_z$). This could also explain the broad hump in the heat capacity around 3.7\,K \cite{Yamashita2011}. 

Measurements of the thermal conductivity, $\kappa$, of \dmit have produced conflicting results. Yamashita \textit{et al.} \cite{Yamashita2010} observe a large residual linear term, attributing this to highly mobile gapless excitations. However, recent studies \cite{BourgeoisHope2019, Ni2019}, find no significant residual linear term as $T \rightarrow 0$ and a much lower value for the thermal conductivity. Yamashita has since published a note \cite{YamashitaNote} showing that they observed both types of behaviour for different samples in their experiment, and proposing that the discrepancy is caused by impurities. However, \citeauthor{BourgeoisHope2019} \cite{BourgeoisHope2019} rejected this explanation.

 In a quasi-1D system, the thermal conductivity is highly dependent on the lattice direction along which it is measured. This is evident in thermal conductivity measurements on \ch{Cs2CuCl4} \cite{Schulze2019}. In the spin liquid phase of \ch{Cs2CuCl4}, a linear contribution, presumably from spinons, is observed in the thermal conductivity along the chain direction, but not  along the other crystal axes. 

This makes sense if we assume that both spinons and phonons can transport heat along the spin chains, but that only phonons can  transfer heat  between the chains. Therefore, if the heat transport has a significant interchain component the thermal resistance adds in series, $1/\kappa_\mathrm{total} \sim 1/\kappa_\mathrm{ph}+1/\kappa_\mathrm{sp}$, where $\kappa_\mathrm{total}$, $\kappa_\mathrm{ph}$, and $\kappa_\mathrm{sp}$ are the total, phonon, and spinon thermal conductivities respectively. This leads to a vanishing thermal conductivity  $\kappa_\mathrm{total}\rightarrow 0$ as $\kappa_\mathrm{ph}\rightarrow 0$ when $T \rightarrow 0$.   When the thermal conductivity is measured in the direction of the chains, the  spinons and phonons are parallel channels and  $\kappa_\mathrm{total}\sim \kappa_\mathrm{ph}+\kappa_\mathrm{sp}$. This leads to two important differences from the series transport scenario: (i) the thermal conductivity should be much larger along the chain directions and (ii) there is a residual linear term as $T \rightarrow 0$ because $\kappa_\mathrm{sp}/T\neq 0$ only for heat transport parallel to the chains. Thus, a prediction of our theory is that there is a linear term, due to spinons, in the thermal conductivity along the chain direction, which alternates between $[1,1,0]$ and $[1,\bar{1},0]$ in different \ch{Pd(dmit)2} layers. 

The direction of thermal transport within the \ch{Pd(dmit)2} planes is not known in most thermal conductivity measurements in \dmit to date. Where the direction has been determined \cite{BourgeoisHope2019}, the heat has always been transported along a crystallographic axis and results in a small thermal conductivity with no linear term -- consistent with our predictions (as this corresponds to the series case). Therefore, all data to date is consistent with our theory. Measurements of thermal conductivity along the $[1,1,0]$ or $[1,\bar{1},0]$ directions would be an important test of our theory as they provide the opportunity to falsify it.

\subsection{Comparison to Previous First-Principles Studies}

It is important to compare our BS-DFT results with other first principles approaches to \dmit. 
The in-plane hopping integrals found by \citeauthor{Nakamura2012} \cite{Nakamura2012} are within the range found from other parametrizations through band structure calculations \cite{Scriven2012,Tsumuraya2013,Jacko2013}. These studies parametrized  monomer or dimer tight-binding models on the basis of band structure calculations; either by fitting to models or via Wannier functions. This approach yields tight-binding models with less anisotropy than our BS-DFT, as summarized in Table \ref{tab:past_dft}, and very weak hopping between the layers \cite{Tsumuraya2013}. 
The  values in Table \ref{tab:past_dft} all lie in or close to the weakly coupled chain regime ($\bar{J}_y^\mathrm{SE}/J_B^\mathrm{SE}\lesssim0.7$;   $\delta{J}^\mathrm{SE}/J_B^\mathrm{SE}\lesssim0.3$).

\begin{table}[h]
	\begin{center}
		\begin{tabular}{c|ccccc}
			Ref. & $t_B$\,(meV) & $t_r$\,(meV) & $t_S$\,(meV) &  $\bar{J}_y^\mathrm{SE}/J_B^\mathrm{SE}$ &  $\delta{J}^\mathrm{SE}/J_B^\mathrm{SE}$\\ 
			\hline 
			\hline
			\cite{Nakamura2012} & 54 & 40 & 45 & 0.62 & 0.14\\ 
			\cite{Jacko2013} & 57 & 40 & 47 &  0.59 & 0.19\\ 
			\cite{Tsumuraya2013}  & 49 & 42 & 37 & 0.65 & 0.16\\
			\cite{Scriven2012} & 49 & 38 & 46 &  0.74 & 0.28\\
		\end{tabular}
		\caption{Previous results for inter-dimer hoppings based on DFT band-structure calculations. To estimate the ratio of exchange couplings, we use the superexchange in the large $U$ limit, 
			$\bar{J}_y^\mathrm{SE}\approx \frac{4}{U}\frac{t_S^2+t_r^2}{2}$,
			$\delta J_y \approx \frac{4}{U}(t_S^2+t_r^2)$
			and $J_B^\mathrm{SE}\approx 4t_B^2/U$, 
			which leads to $J_B^\mathrm{SE}/\bar{J}_y^\mathrm{SE}\sim 2t_B^2/(t_r^2 + t_S^2)$. 
			All of these models lie in or close to a weakly coupled chain regime ($\bar{J}_y^\mathrm{SE}/J_B^\mathrm{SE}\lesssim0.7$;   $\delta{J}^\mathrm{SE}/J_B^\mathrm{SE}\lesssim0.3$) which gives rise to a gapless spin-liquid state \cite{Zheng1999, Zheng2007, Yunoki2006, Hayashi2007, Tocchio2014, Ghorbani2016}.}\label{tab:past_dft}
	\end{center} 
\end{table}

Furthermore, \citeauthor{Nakamura2012}'s constrained RPA calculations of the Coulomb interactions allow for a fairly direct comparison with our BS-DFT results. We find that our values for $U\!-\!V_{ij}$ are similar in magnitude, following the same trend, and we also find a similar correlation strength, $U/t_B\approx 7$. In our notation, \citeauthor{Nakamura2012} found that $J_B=262$\,K, $\bar{J}_y=157$\,K, $\delta{J_y}=47$\,K, and $K_4=43$\,K. As $\bar{J}_y/J_B=0.60<0.7$ and  $\delta{J_y}/J_B = 0.18<0.3$ the CRPA approach should still be reasonable, although \citeauthor{Nakamura2012}'s values are less quasi-one-dimensional than those found in our BS-DFT calculations. We have performed BS-DFT calculations using pure funtionals (LDA and PBE) and find that these provide a poor description of \dmit, as witnessed by a large spin contamination. This may explain the discrepancy. Furthermore, BS-DFT calculations are differences of total energies, whereas band structure calculations are based on Kohn-Sham eigenvalues. The former are far more accurate in DFT.
Regarding the interlayer hopping integral, \citeauthor{Nakamura2012} do not report a value, but similar calculations by \citeauthor{Tsumuraya2013} find that there is very weak hopping between the layers  \cite{Tsumuraya2013}. Thus it is safe to assume that $J_z<18$\,K. Whence, $2K_4>|\delta J_y|\!+\!|J_z|$ and \citeauthor{Nakamura2012}'s parameters also place \dmit in the quantum disordered regime. 
Moreover, a recalculation of Fig. \ref{fig:Tn_plot} with Nakamura \textit{et al}'s values for $\bar{J_y}$, $K_4^\prime$, and $K_4^{\prime\prime}$ (see the Supplementary Material \cite{Sup}), is a reproduction of Fig. \ref{fig:Tn_plot} because $T_N$ depends only on $\delta J_y$, $J_z$, and $K_4$.

The quasi-one-dimensionality of the spin Hamiltonian derived from band structure calculations has been previously noted \cite{Nakamura2012, Tsumuraya2013}, although detailed many-body calculations have not previously been performed in this regime of the scalene triangular lattice. 

\section{Conclusions}

We have used an atomistic approach to parametrize an extended Hubbard model, and thence a spin model, for the spin-liquid candidate \dmit. This revealed a frustrated scalene triangular lattice where the largest coupling along the stacking direction is nearly three times larger than the others. We showed that, in the quasi-one-dimensional limit relevant to \dmit, the difference in the interchain coupling acts identically to an unfrustrated interchain coupling and favors long-range magnetic order. This interaction competes with ring exchange, which promotes quantum disorder. Our DFT calculations show that, in \dmit, $2K_4>|\delta J|+|J_z|$ and we therefore predict that \dmit does not order magnetically even at $T=0$. Thus, we propose that the `spin-liquid' behaviour is a remnant of TLL behavior in weakly coupled 1D spin chains.

We predict that thermal transport along the chains is very different from that with a significant transverse component. This is because in the former case spinons and phonons act as parallel channels whereas in the latter they transport heat in series. This means (i) that the thermal conductivity along the chains is much larger than the thermal conductivity across the chains and (ii) that there is a large linear contribution to the thermal conductivity along the chains (due to the spinons), but no linear contribution to transverse thermal conductivity. Behaviours consistent with both the parallel and series pictures have been observed experimentally, resulting in some controversy. Where the thermal transport direction is know it is along crystallographic axes and consistent with our predictions for series heat transport. Measurements of heat transport along the chains (in the $[1,1,0]$ and $[1,\bar{1},0]$ directions) would provide a key test of our theory and could either confirm or falsify it.

\begin{acknowledgments}
We thank Amie Khosla and Ross McKenzie for helpful conversations. 
M\'esocentre of Aix-Marseille Universit\'e is acknowledged for allocated HPC resources.
This work was supported by the Australian Research Council through Grants No. DP160100060 and DP180101483.
\end{acknowledgments}

\bibliographystyle{apsrev4-1}

\clearpage
\section*{Supplementary Material}

\subsection*{CRPA analysis using Hamiltonian parameters from Nakamura \textit{et al.}}

\begin{figure}[h!]
	\centering
	\includegraphics[width=8cm]{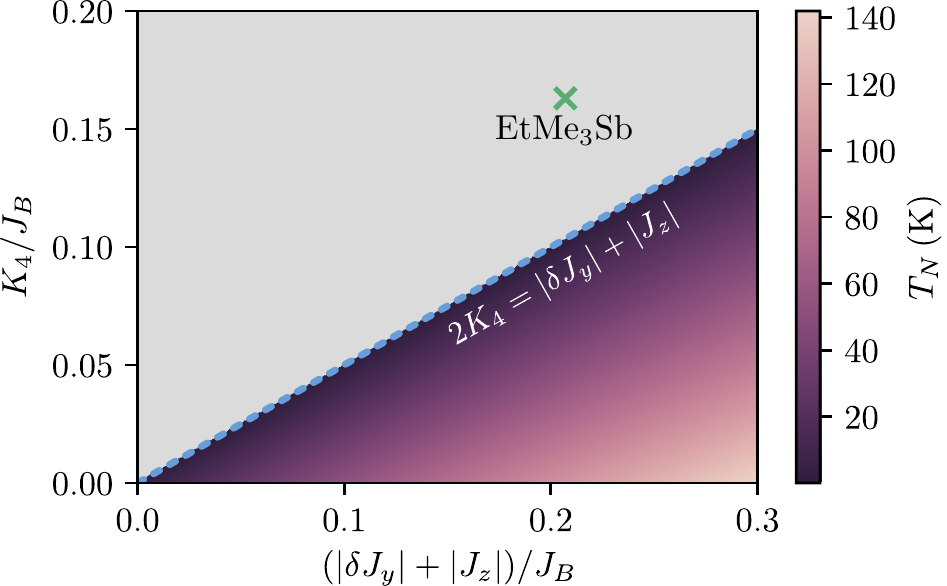}
	\caption{A recalculation of Fig. \ref{fig:Tn_plot} in the main text with Nakamura \textit{et al}'s \cite{Nakamura2012} values for $\bar{J_y}$, $K_4^\prime$, and $K_4^{\prime\prime}$ (shown in Table \ref{tab:Nakamura}). The green cross indicates their values for $K_4$ and $\delta J_y$, which lie in the disordered regime. It is otherwise indistinguishable from Fig. \ref{fig:Tn_plot} because $T_N$ depends only on $\delta J_y$, $J_z$, and $K_4$.}
	\label{fig:Tn_plot_nakamura}
\end{figure}

\citeauthor{Nakamura2012} \cite{Nakamura2012} previously parametrized the extended Hubbard model for \dmit using the constrained random-phase approximation (RPA) and maximally localized Wannier orbitals. Their values are shown in Table \ref{tab:Nakamura}, which lead to $K_4=43$\,K,  $K_4^\prime=63$\,K, and  $K_4^{\prime\prime}=78$\,K. Fig. \ref{fig:Tn_plot_nakamura} is equivalent to Fig. \ref{fig:Tn_plot} in the main text, but it is based on their values. We find that their parametrization leads to the same prediction that ring exchange overcomes the unfrustrated couplings and causes spin disorder, since $2K_4>|\delta J_y|$. Note that  \citeauthor{Nakamura2012} \cite{Nakamura2012} did not calculate an inter-layer coupling, $J_z$, but another study using similar methods finds very weak hopping between the layers \cite{Tsumuraya2013}.

\begin{table}[h!]
	\begin{center}
		\begin{tabular}{c|ccccc}
			$ij$& $J_{ij}$\,(K) & $J_{ij}^\mathrm{SE}$\,(K) & $J_{ij}^\mathrm{DE}$\,(K) 
			& $t_{ij}$\,(meV) & $U\!-\!V_{ij}$\,(meV)\\ 
			\hline 
			\hline
			B & 262 & 335 & -73 & 54 & 410\\ 
			r & 133 & 170 & -37 & 40 & 440\\
			S & 180 & 240 & -60 & 45 & 390\\
		\end{tabular}
		\caption{Nearest neighbour interactions between dimers of \dmit calculated by Nakamura \textit{et al.} \cite{Nakamura2012}. Labels as shown in Figure \ref{fig:geometry}a of the main text. The full Heisenberg exchange ($J_{ij}$) is a sum of the superexchange ($J_\mathrm{SE}$) and the direct exchange ($J_\mathrm{DE}$). $t_{ij}$ is the effective hopping between dimers and $U\!-\!V_{ij}$ is the effective Coulomb interaction on each dimer.}\label{tab:Nakamura}
	\end{center} 
\end{table}

\newpage
\subsection*{Effect of DFT functional on BS-DFT coupling parameters}

The magnetic exchange coupling calculated with broken-symmetry density functional theory (BS-DFT) is strongly dependent on the amount of Hartree-Fock exchange (\%HFX) used within the chosen functional.
This feature has been widely studied in the molecular magnet framework \cite{Martin1997, Sorkin2008, Reiher2001, Swart2004, Dai2005, Lawson2005, Pierloot2006, Rong2006, Brewer2006, Vargas2006, Ovcharenko2007, Valero2008} and its influence on the model Hamiltonian parameters (Heisenberg and Hubbard) investigated by some of the authors \cite{David2017}.   
Table \ref{tab:dft_hfx} shows a comparison of some Hamiltonian parameters (c.f. Table \ref{tab:dft} in the manuscript) calculated using B3LYP with 20 and 50\%HFX.
In the manuscript, we used the original B3LYP, which has 20\%HFX.
As shown in the table \ref{tab:dft_hfx} this variation mainly affects the Hubbard $U$ parameter, which evolves proportionally with the amount of HFX. 
Indeed increasing the amount of HFX tends to reduce the self interaction error yielding molecular orbitals less delocalized.
Therefore the on-site repulsion energy $U$ becomes more important.
The kinetic exchange contribution $J_{ij}^\mathrm{SE}$, being proportional to $1/U$, is dramatically reduced for the amount of HFX going from 20\% to 50\%.
Therefore the total magnetic coupling $J_{ij}$ appears weaker for 50\%HFX than 20\%HFX.

\begin{table}[h!]
	\begin{center}
		\begin{tabular}{c|cc|cc}
			& \multicolumn{2}{c}{$ij=$r} & \multicolumn{2}{c}{$ij=$S}\\
			\hline
			&20\%HFX & 50\%HFX & 20\%HFX & 50\%HFX\\
			\hline
			$J_{ij}$\,(K) & 146 & 38 & 124 & 52\\
			$J_{ij}^\mathrm{SE}$\,(K) & 168 & 53 & 168 & 101\\ 
			$J_{ij}^\mathrm{DE}$\,(K)  & -22 & -15 & -44 & -49\\ 
			$t_{ij}$\,(meV) & 47 & 43 & 44 & 54\\ 
			$U\!-\!V_{ij}$\,(meV) & 600 & 1627 & 519 & 1326\\ 
			\hline 
		\end{tabular}
		\caption{Comparison of BS-DFT results using B3LYP with 20\% HFX and 50\% HFX.}\label{tab:dft_hfx}
	\end{center} 
\end{table}

\end{document}